\documentclass[prapplied,longbibliography,twocolumn,superscriptaddress,aps,amssymb]{revtex4-2} 
\usepackage{amsmath, verbatim, graphicx, fancyhdr, color}
\usepackage{float}
\usepackage{braket}

\usepackage[sort&compress]{natbib}
\usepackage{wrapfig}
\usepackage{siunitx}
\DeclareSIUnit{\rpm}{r.p.m.}
\usepackage[dvipsnames]{xcolor}
\usepackage{setspace}
\usepackage{titlesec}

\date{ {\normalfont \today}}

\begin{document}
\title{ A robust tip-less positioning device for near-field investigations: \\Press and Roll Scan (\textit{PROscan}) }
\author{Hsuan-Wei Liu}
\thanks{These two authors contributed equally }
\affiliation{Max Planck Institute for the Science of Light, D-91058 Erlangen, Germany}
\affiliation{Friedrich-Alexander-Universit\"at Erlangen-N\"urnberg, Dept. Physics, D-91058 Erlangen, Germany}

\author{Michael A. Becker}
\thanks{These two authors contributed equally }
\affiliation{Max Planck Institute for the Science of Light, D-91058 Erlangen, Germany}

\author{Korenobu Matsuzaki}
\affiliation{Max Planck Institute for the Science of Light, D-91058 Erlangen, Germany}

\author{Randhir Kumar}
\affiliation{Max Planck Institute for the Science of Light, D-91058 Erlangen, Germany}

\author{Stephan G\"{o}tzinger}
\affiliation{Friedrich-Alexander-Universit\"at Erlangen-N\"urnberg, Dept. Physics, D-91058 Erlangen, Germany}
\affiliation{Max Planck Institute for the Science of Light, D-91058 Erlangen, Germany}
\affiliation{Friedrich-Alexander-Universit\"at Erlangen-N\"urnberg, Graduate School in Advanced Optical Technologies (SAOT), D-91052 Erlangen, Germany}

\author{Vahid Sandoghdar}
\affiliation{Max Planck Institute for the Science of Light, D-91058 Erlangen, Germany}
\affiliation{Friedrich-Alexander-Universit\"at Erlangen-N\"urnberg, Dept. Physics, D-91058 Erlangen, Germany}

\begin{abstract}
Scanning probe microscopes scan and manipulate a sharp tip in the immediate vicinity of a sample surface. The limited bandwidth of the feedback mechanism used for stabilizing the separation between the tip and the sample makes the fragile nanoscopic tip very susceptible to mechanical instabilities. We propose, demonstrate and characterize a new alternative device based on bulging a thin substrate against a second substrate and rolling them with respect each other. We showcase the power of this method by placing gold nanoparticles and semiconductor quantum dots on the two opposite substrates and positioning them with nanometer precision to enhance the fluorescence intensity and emission rate. We exhibit the passive mechanical stability of the system over more than one hour. The design concept presented in this work holds promise in a variety of other contexts, where nanoscopic features have to be positioned and kept near contact with each other.

\end{abstract}

\maketitle

\section{Introduction}

Since the invention of the scanning tunneling microscope (STM) in 1981 \cite{binnig1982tunneling}, scanning probe microscopy (SPM) has become indispensable in nanoscience and surface science, where structures below the optical diffraction limit are investigated down to individual atoms. Two of the most prominent SPM methods that followed STM are atomic force microscopy (AFM) \cite{binnig1986atomic} and scanning near-field optical microscopy (SNOM) \cite{lewis1984development,pohl1984optical}. The central and common feature of all these methods is a sharp tip that is operated in the immediate vicinity of a sample surface. 

In addition to high-resolution imaging, SPMs have also proven very valuable for manipulation and control of nanoscopic interactions. A leading example in nano-optical studies has been the placement of nano-antenna structures at the end of sharp dielectric tips so as to couple them to emitters in a controlled fashion \cite{kalkbrenner2001single,farahani2005single, anger2006enhancement, kuhn2006enhancement,kuhn2008modification, matsuzaki2017strong,singh2018nanoscale,gross2018near}. Similar experiments have also been reported on gap plasmons generated by coupling an emitter to the junction between a metallic tip and a metalized substrate \cite{luo2019electrically, kuhnke2017atomic,park2018radiative,park2019tip,yang2020sub}. However, these efforts remain very challenging and often not accessible to the wider use.

Aside from the difficulties encountered in the fabrication and handling of tips, an important hurdle in SPM-based efforts is the low bandwidth of the feedback signal (e.g., via tunneling current) used to maintain the tip-sample separation at a nanoscopic value. This makes SPMs highly susceptible to mechanical perturbations and irreversible tip damage. In shear-force feedback, which is commonly used in SNOM \cite{karrai2000interfacial}, it is particularly difficult to control distances with 1\,nm precision although theoretical calculations predict optimal plasmonic enhancements at near-to-contact distances \cite{kongsuwan2018suppressed}. In this work, we present a tip-less platform for performing mechanically robust and controllable experiments deep in the optical near field.

\begin{figure}[ht!]
	\begin{center}
	\includegraphics[width=0.48\textwidth]{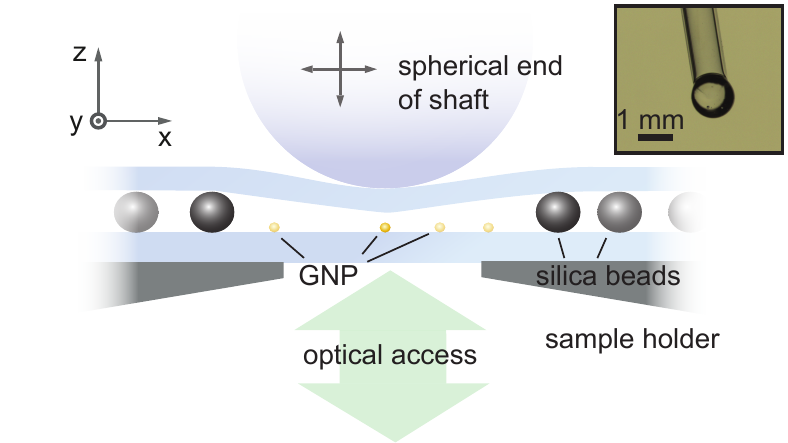}
	\caption{Schematic illustration of a PROscan device. A substrate is pressed and rolled against a second one, carrying gold nanoparticles (GNP) as nano-antenna. The medium to be studied is prepared on the upper substrate. \textit{Inset:} Glass capillary with a spherical end, molten by a CO$_2$ laser, is used as a handle to press and roll the top substrate.}
\vspace{-20pt}
\label{fig:setup}
	\end{center}
\end{figure}

\section{PROscan concept and its realization scheme}

\begin{figure*}[ht!]
	\includegraphics[width=\textwidth]{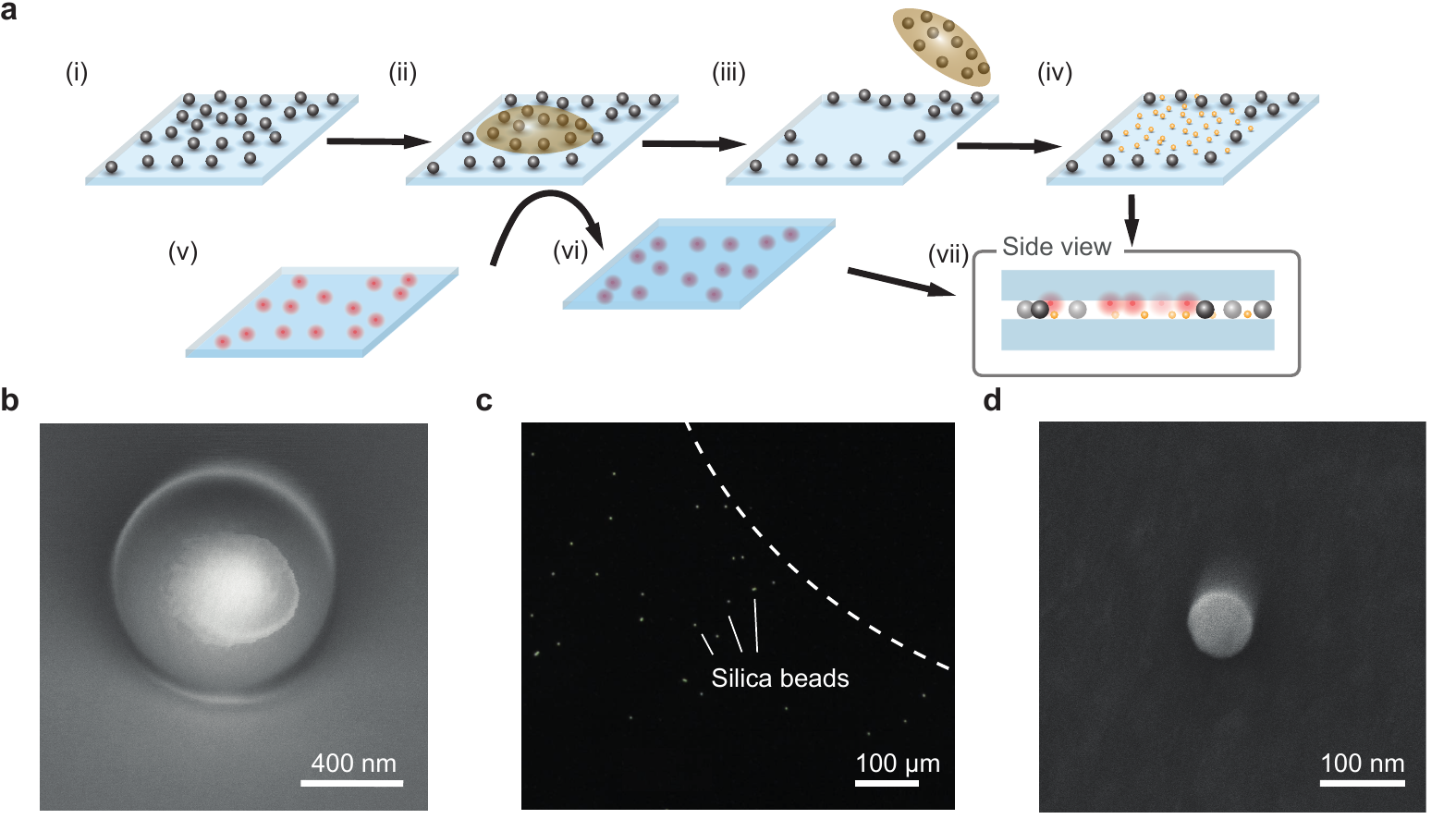}
	\caption{\textbf{a}, Sample preparation procedure: i) Silica beads are spin coated onto a clean (bottom) glass substrate. A cleaning polymer is applied to the central region (ii) and is removed after drying (iii). iv) Nanoprobes are spin coated onto the same substrate. (v) A solution of emitters is spin coated onto a second (top) substrate. The top substrate is flipped (vi) and sandwiched onto the bottom substrate (vii).
\textbf{b}, Scanning electron micrograph of a single silica bead with a diameter of \SI{800}{\nano\meter} coated with chromium. \textbf{c}, Bottom substrate after removal of the cleaning polymer. The dashed white line marks the boundary between the region with dispersed silica beads and the cleaned area.  \textbf{d}, Helium-ion microscope image of an individual gold nanoparticle.}
	\label{fig:rolling mechanism}
\vspace{-10pt}
\end{figure*}

Figure\,\ref{fig:setup} sketches the heart of the new device. The main strategy is to bulge a substrate by a very small amount towards a second substrate that is designed to be much less flexible. In our current work, we chose the flexible substrate to be a cover glass of thickness \SI{100}{\micro\meter}, which was thoroughly cleaned with de-ionized water and non-halogenated solvents (acetone and isopropanol). We used a conventional microscope cover glass of thickness \SI{170}{\micro\meter} as the lower substrate and supported it with a steel bracket that contained a small central opening of \SI{5}{mm} to allow for optical detection through an oil-immersion microscope objective (Olympus UPlanSApo $100\times$, NA $=1.4$). The substrate was treated with oxygen plasma to remove organic residuals and to make its surface hydrophilic.  

To produce a miniaturized handle for exerting local pressure on the upper substrate, we melted the end of a glass capillary to a quasi-spherical shape of diameter $\sim \SI{1}{\milli\meter}$ (see inset in Fig.\,\ref{fig:setup}). The capillary shaft was mounted to a piezo-electric element (Piezosystem Jena, Tritor 38) and was used to press on the  substrate. A small amount of epoxy was applied to the spherical end to enhance surface friction. As we shall see below, a typical voltage of few tens of volts are sufficient to bend the upper cover glass by several 100\,nm over a lateral span of about 10\,mm. 

By placing spherical spacers between the two substrates, we adjust their separation to the desired range and allow for rolling the entire upper substrate. Figure\,\ref{fig:rolling mechanism}a illustrates the fabrication process. We first spin coated silica beads (Bangs Laboratories Inc.) on the supporting lower substrates (\SI{30}{\second} at a speed of \SI{3000}{\rpm}) to yield an average coverage of one particle per $100\times100\,\mu m^2$. In our current report, we varied the silica bead diameter between $0.8-3$ \si{\micro\meter}. Figure\,\ref{fig:rolling mechanism}b presents an electron microscope (EM) micrograph of such a bead. The particles are then removed from the central region of the substrate by an adhesive cleaning polymer (First Contact, Photonic Cleaning Technologies). Here, we first apply a solution of \SI{100}{\micro\liter} of the cleaning polymer to the center of the bottom substrate to form a droplet spread over roughly 5-10\,mm. After drying, the polymer is simply peeled off to remove the silica beads that were in contact with it. This process leaves the surface in a very clean state. Nevertheless, we expose the assembly to 5 minutes of $\rm O_{2}$ plasma in order to eliminate any nanoscopic residual amount of polymer. A dark-field optical microscope image of the outcome is shown in Fig.\,\ref{fig:rolling mechanism}c. The clean central area allows one to press the sample in the axial direction until the two glass substrates touch. 

To mimic the sharp end of an SPM tip, one of the substrates is decorated with a well-defined ``nanoprobe'' of choice. In our current work, we demonstrate the principle of this step with gold nanoparticles (GNP) of diameter \num{80} \si{\nano\meter}, which we spin coat on the lower substrate for \SI{30}{\second} at a speed of \SI{3000}{\rpm} (see step iv in Fig.\,\ref{fig:rolling mechanism}a). We chose a GNP coverage corresponding to an average particle separation of \SI{10}{\micro\meter} to facilitate diffraction-limited optical detection in an uncrowded region. Figure\,\ref{fig:rolling mechanism}d shows a helium-ion microscope image of a GNP. 

The medium of interest to be studied, e.g., thin films, nanoparticles, or molecules, is placed on the substrate opposite to the one containing the nanoprobe. In this work, we used CdSe/CdS core/shell colloidal quantum dots (qdot) with a core size of \SI{4}{\nano\meter} and a total diameter of \SI{15}{\nano\meter} \cite{orfield2016quantum,matsuzaki2017strong}. We sparsely deposited the qdots onto the upper glass substrate. To do this, a toluene solution of the colloidal qdots was diluted to nanomolar concentrations. \SI{20}{\micro\liter} of this suspension was spin coated in a two-step process (\SI{1000}{\rpm} for \SI{30}{\second} followed by \SI{3000}{\rpm} for \SI{3}{\second}). Subsequently, the upper substrate was flipped and placed on top of the bottom substrate (see step vii in Fig.\,\ref{fig:rolling mechanism}a). Various modes of optical microscopy (dark-field, iSCAT, fluorescence, etc.) can be employed to identify the individual nanoprobes. 

As we demonstrate below, the upper substrate can be pressed and rolled to scan a qdot against a GNP with nanoscopic precision (see also Fig.\,\ref{fig:setup}). We thus refer to this technique as \textit{PROscan}. We emphasize that the choices of the two substrates (e.g., material, thickness), rolling mechanism (e.g., choice of beads, nano-lubrication), and the nanoprobe can be varied and optimized for different applications. In particular, one could use bottom-up (e.g., self-assembly) or top-down (e.g., electron beam lithography) fabrication for the realization of various nanoprobes such as cones \cite{hoffmann2015fabrication, matsuzaki2017strong}.

\subsection{Lateral scan and position control}
\begin{figure}[h!]
	\centering
	\includegraphics[width=0.48\textwidth]{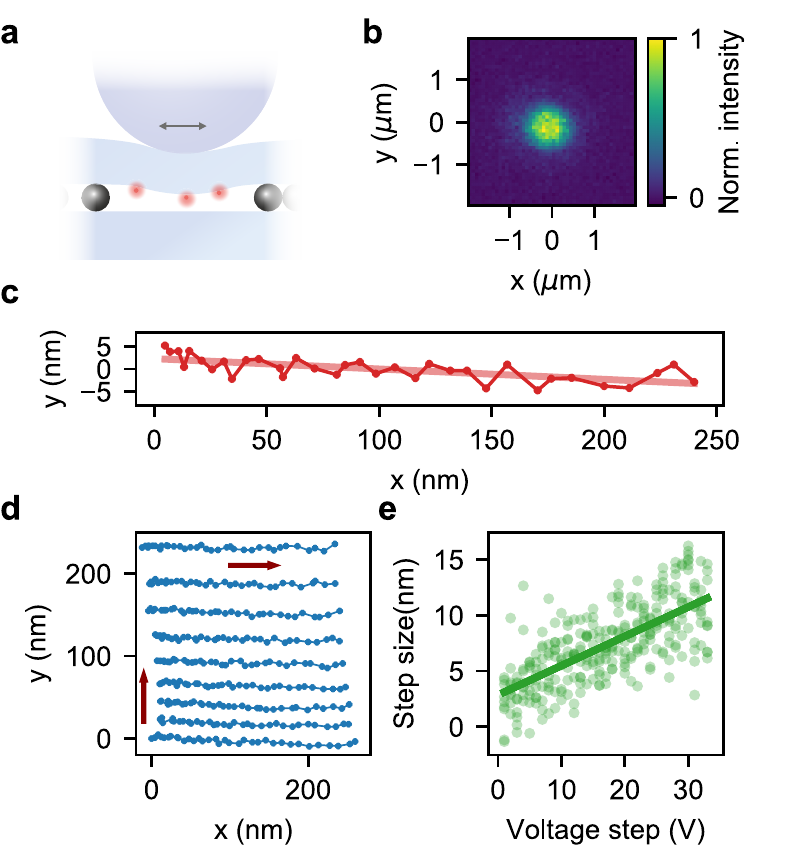}
	\caption{\textbf{a}, Schematic of the measurement configuartion used for characterizing lateral scans. \textbf{b}, Fluorescence image of an individual qdot placed on the top substrate. Localization of the qdot in images recorded at each location during the rolling and scanning process report on the trajectory of the top substrate . \textbf{c}, An exemplary trajectory of a qdot during an $x$-scan of the top substrate. The bright red line depicts a linear fit to the data. \textbf{d}, Example of the qdot trajectories recorded during a $xy$-scan of the top substrate. The directions of the 2D scan are indicated by the red arrows. \textbf{e}, Variations of the measured step size for the applied voltage steps in $x$ direction for all lateral scans shown in \textbf{d}. Solid green line represents a linear fit to the data.}
	\label{fig:horizontal}
\end{figure}

PROscan allows one to approach a selected nanoprobe both laterally and axially to the location of the sample under investigation. A range of signals can be used to monitor or control the separation of the two substrates during this process. In this work, we used a combination of interferometry, modification of the GNP plasmon resonance, and fluorescence enhancement of qdots. To examine the quality of the lateral position control, we recorded the trajectories of single qdots while rolling the upper substrate against a naked (i.e., without any GNP) lower substrate (see Fig.\,\ref{fig:horizontal}a). The qdots were excited in a wide-field arrangement with a continuous-wave laser beam at a wavelength of \num{532}\,\si{\nano\m} focused on the back focal plane of the microscope objective. A fluorescence image was recorded at each lateral \textit{xy}-scan, and the point-spread function (PSF) of the qdot (see Fig.\,\ref{fig:horizontal}b) was analyzed by fitting a two-dimensional Gaussian function. The high signal-to-noise ratio (SNR) allowed us to localize a qdot \cite{weisenburger2015light} with an average precision of \SI{2.4}{\nano\meter} in both directions.

Figure\,\ref{fig:horizontal}c displays an example of a trajectory while a linear voltage ramp of \SI{1}{\volt} was applied to the piezo element along the \textit{x} direction without the use of any feedback control. We fit the data with a linear function and obtain a slope of $m = -0.023\pm 0.005$, corresponding to a tilt angle of \SI{1.3\pm0.3}{\degree}. In Fig.\,\ref{fig:horizontal}d, we present a series of \textit{x} scans recorded at different \textit{y} locations. Here, we observe an average tilt angle of  \SI{1.5}{\degree} with a standard variation of \SI{0.5}{\degree}. Such a small tilt could be caused by a slight misalignment of the piezo element with respect to the camera, but the variations among the individual scans lead us to attribute the observed variation to a small cross talk between the two axes of the piezo. Moreover, we note that the step size increases at higher applied voltages while it is smaller again after the turning points in a zig-zag pattern. When we scan only in one direction, however, the step size becomes uniform after a while (see Fig.\,\ref{fig:qdot}d). Figure\,\ref{fig:horizontal}e displays the measured step sizes as a function of the applied voltage. We find an average step size of \SI{7.3}{\nano\meter} with a standard deviation of \SI{3.5}{\nano\meter} in the scan direction and a jitter of \SI{2.74}{\nano\meter}.

In conventional SPM, the relative lateral position of the tip and the sample is usually passively scanned by piezo-electric elements although actively stabilized scanners have also become common. The finesse of the scanning grid depends on the application and can be as small as \AA ngstroms. The same instrumentation can also be used for applications where the probe is used to manipulate the sample, e.g., in AFM lithography \cite{yasutake1993modification,snow1994fabrication} or plasmonic nano-antennas \cite{kalkbrenner2001single,farahani2005single, anger2006enhancement, kuhn2006enhancement,kuhn2008modification, matsuzaki2017strong,singh2018nanoscale,gross2018near, luo2019electrically, kuhnke2017atomic,park2018radiative,park2019tip,yang2020sub}. In the latter case, however, a passive pre-determined knowledge of the tip-sample position is not a requirement. It would be sufficient to measure and monitor the relative position during the experiment. The data in Fig.\,\ref{fig:horizontal}c,d show that although the lateral scans in this very simple implementation of PROscan are not as uniform as in conventional SPM, they do achieve nanometer precision. This performance can be further improved by employing additional feedback mechanisms, e.g., by measuring the actual position of an emitter and correcting for small step size errors. 

\subsection{Axial position control}
The most crucial feature of an SPM technique is to approach a nanoprobe to a sample with nanometer or sub-nanometer precision. In practice, one usually reduces the distance with a translation stage in an iterative manner until the characteristic near-field signal is detected. In STM, one uses the tunneling current as a measure for the sample-probe distance, whereas the shear-force signal is used in SNOM \cite{karrai2000interfacial,betzig1991breaking}. In PROscan, one can use optical microscopy to determine and monitor the position of the upper substrate with nanometer precision. 

\subsubsection{Coarse approach} 
\begin{figure}[ht!]
	\centering
	\includegraphics[width=0.48\textwidth]{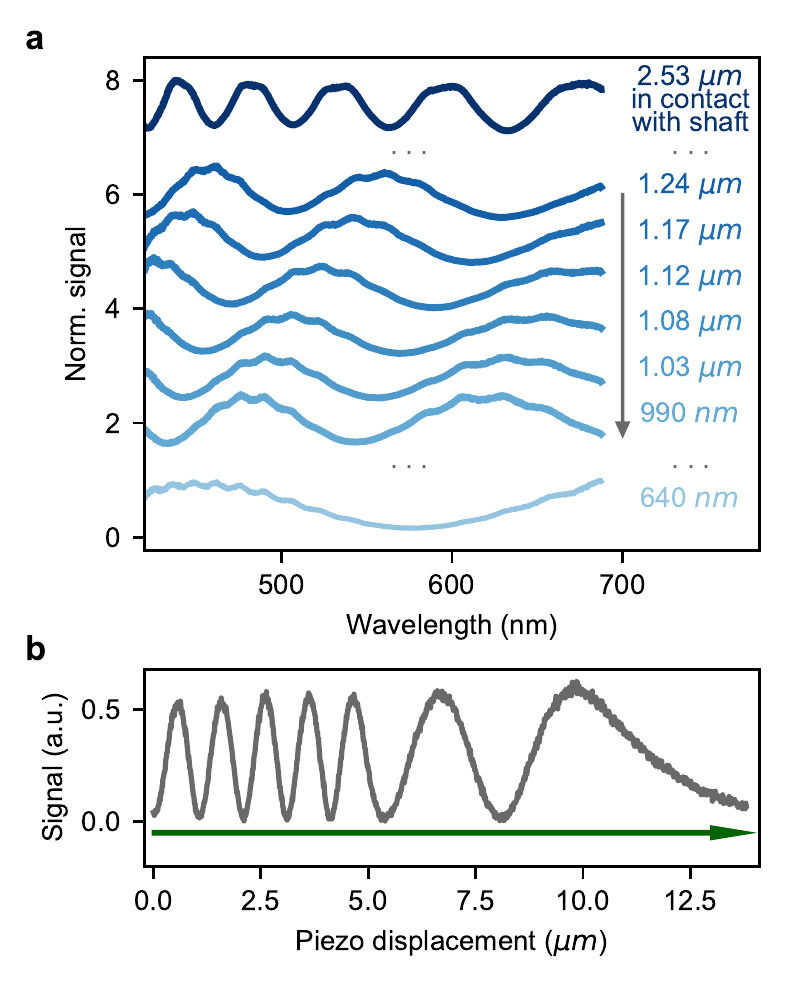}
	\caption{\textbf{a}, Interference spectra for different gap sizes between the two glass substrates using white-light illumination. The dark blue spectrum on top was recorded after the capillary shaft contacted the top substrate. \textbf{b}, Reflected intensity of a laser beam at $\lambda = \SI{532}{\nano\meter}$ as the upper substrate is lowered further. A piezo displacement of about 13\,$\mu$m leads to a substrate displacement of about 1.9\,$\mu$m.}
	\label{fig:interference}
\end{figure}

For gaps larger than the wavelength of visible light, interference fringes can be used to deduce the distance between the two substrates. As shown in Fig.\,\ref{fig:interference}a, we use a white-light excitation source to record the interference signal as a function of wavelength at each axial position. Analysis of this information allows us to deduce an absolute distance $d$ between the two substrates according to the formula $d = \frac{\lambda_1\cdot \lambda_2}{2(\lambda_2-\lambda_1)}$ obtained from the free spectral range of an optical cavity formed in air. Here, $\lambda_1$ and $\lambda_2$ are the wavelengths of neighboring intensity maxima. This simple strategy ceases to work at small separations, where one no longer records full oscillations (see lower trace in Fig.\,\ref{fig:interference}a).

Hence, we augment our knowledge of the substrate separation by measuring the relative displacement using monochromatic interferometry ($\lambda=\SI{532}{\nano\meter}$). As shown in Fig.\,\ref{fig:interference}b, each oscillation indicates a displacement of $\lambda /2 = \SI{266}{\nano\meter}$. The monotonic change in the periodicity as a function of the applied piezo voltage indicates the reduction in the bending capability of the substrate. For this measurement we used \SI{3}{\micro\meter}-sized silica beads to start with a larger separation between the substrates.

\begin{figure*}[ht!]
	\includegraphics[width=\textwidth]{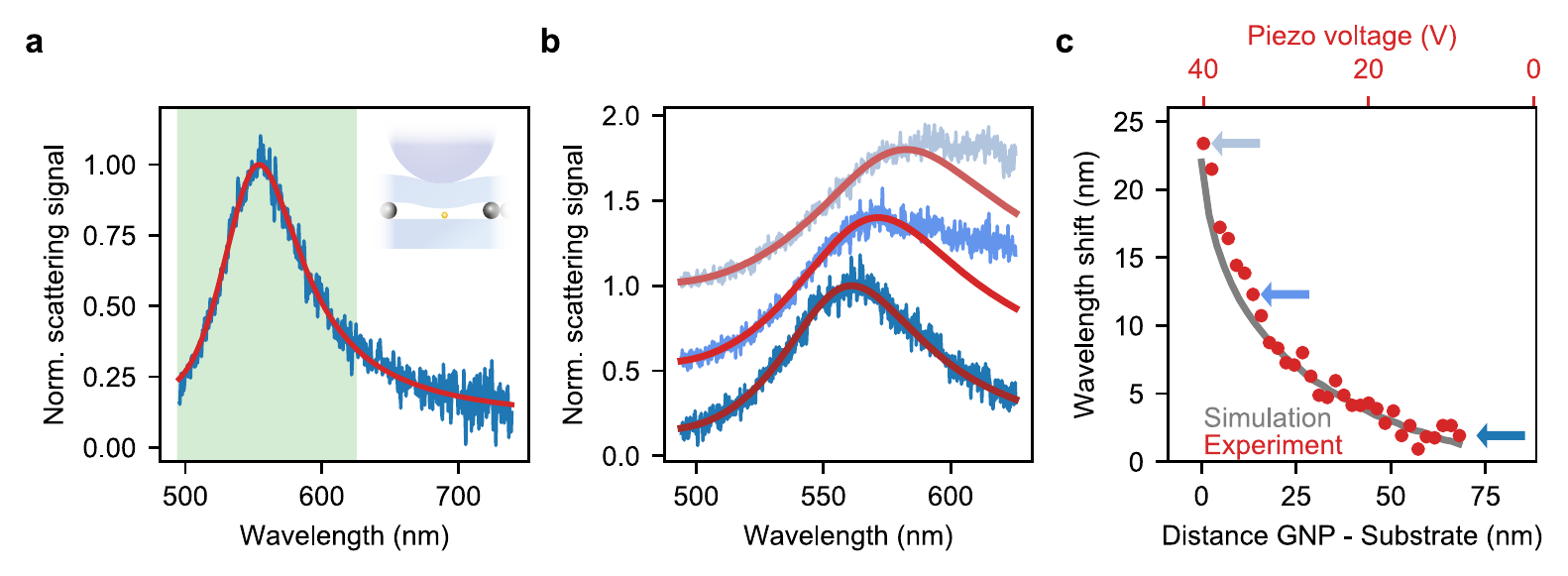}
	\caption{\textbf{a}, Plasmon spectrum of an individual GNP on a glass substrate recorded in a dark-field arrangement. The red solid line is a fit to the data using a theoretical model. The green highlighted area indicates the measurement region in \textbf{b}. \textit{Inset}: Measurement configuration showing the top glass substrate being lowered to the individual GNP placed on the lower substrate. \textbf{b}, Normalized plasmon spectra recorded at three different axial positions as marked by the correspondingly color-coded arrows in (c). Solid curves display fits (red). \textbf{c}, Measured wavelength shifts (red symbols) extracted from plasmon spectra recorded at different axial positions as a function of the applied piezo voltage (upper horizontal axis). Grey solid curve overlays the theoretical resonance shift calculated for different gap distances (lower horizontal axis).  Blue color-coded arrows indicate data points from the spectra shown in \textbf{b}.}
	\label{fig:vertical}
\end{figure*}

\subsubsection{Fine approach}The interferometric measurements discussed above cannot report on the separation between a nanoscopic probe such as a GNP and the substrate on the opposite side. To demonstrate that we can control this distance with nanometer precision in the near field, we measured the modification of the plasmon resonance of a single GNP \cite{buchler2005measuring,hakanson2008coupling} placed on the bottom substrate, while lowering the upper glass substrate. Figure\,\ref{fig:vertical}a displays an example of the plasmon spectrum of a single \num{80}\,\si{\nano\meter} GNP when the second substrate is still far apart.  In this unperturbed case, the resonance maximum was at about \SI{560}{\nano\meter}. The measurement was performed in a dark-field configuration, where $p$-polarized white light (Energetiq EQ-99 LDLS) was focused into the back focal plane of the microscope objective in a total internal reflection (TIR) configuration. In this manner, a plasmon mode that is polarized orthogonal to the glass substrate was excited. The scattered light from the particle was collected by the same objective, and the reflected light was blocked in the Fourier plane by the edge of a razor blade. Here, one needs to account for changes of both the excitation spectrum and the scattering background caused by the bending of the substrates and by spurious interference effects. To improve the SNR, we applied a moving-average filter to the data and calculated the mean over 3 consecutive steps.

In Fig.\,\ref{fig:vertical}b, we present three examples of plasmon spectra recorded at different separations of the upper substrate. The displayed wavelength range corresponds to the green shaded area in Fig.\,\ref{fig:vertical}a. To determine the resonance wavelength, we fit the normalized scattering spectrum to its theoretical counterpart, taking into account the dipolar approximation and an effective polarizability \cite{hakanson2008coupling,kalkbrenner2004tomographic}. At higher wavelengths, we observe slight deviations from the theoretical line shape, which could be caused by temporary near-field charging effects \cite{zapata2016plasmon,collins2016single}. We, thus, only consider the region up to \SI{590}{\nano\meter} in fitting the spectrum (see red curve). Overall, the spectra clearly report on the expected near-field change in the resonance frequency and linewidth \cite{buchler2005measuring,hakanson2008coupling}, which can be intuitively understood as the result of the interaction between the dipole moment associated with the plasmon mode and its mirror image in the glass substrate. 

The symbols in Fig.\,\ref{fig:vertical}c plot the extracted plasmon resonance as a function of the applied piezo voltage in the $z$ direction. The color-coded arrows indicate the corresponding spectra in Fig.\,\ref{fig:vertical}b. We also present the results of numerical FDTD simulations for the resonance shift of the system. Although the measured plasmon spectra cannot be exactly matched to the simulated spectra, the wavelength shift shows a very good agreement with the measurement. This agreement shows that we achieve nanometer control of the axial position. 

\begin{figure*}[t!]
\centering
\includegraphics[width=\textwidth]{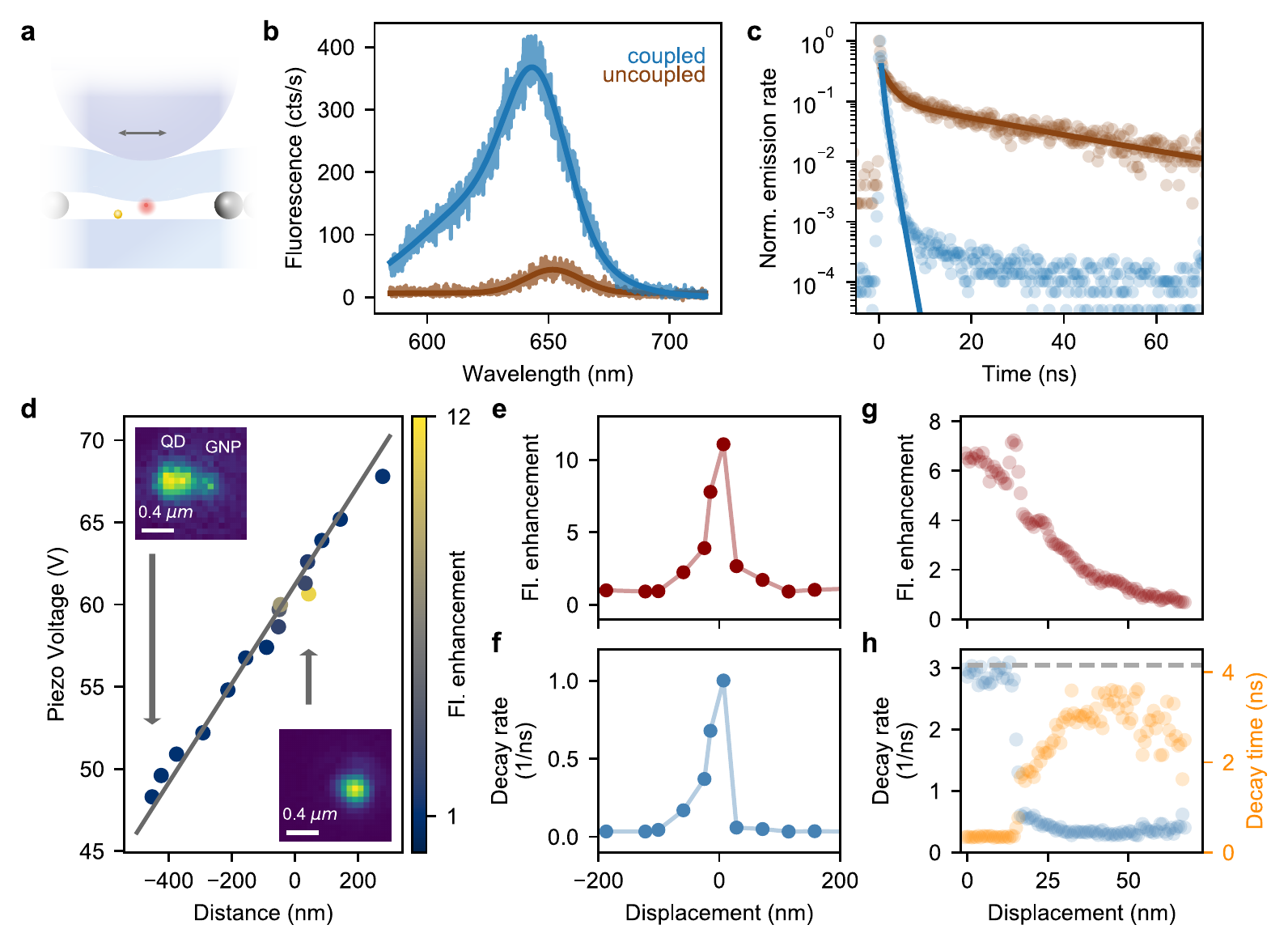}
\caption{\textbf{a}, Schematic of measurement configuration: A single qdot is scanned over an individual GNP. \textbf{b}, Fluorescence spectra of a single CdSe/CdS qdot before (brown) and after coupling (blue) to an individual gold nanoparticle. \textbf{c}, Fluorescence decay time measurement of the uncoupled (brown) and coupled (blue) qdot. Both curves are fitted with a bi-exponential decay function (solid lines). The slow part of the data for the coupled case was not taken into account since it corresponds to a drop of three orders of magnitude in the signal. \textbf{d}, Applied piezo voltage as a function of the lateral GNP-qdot distance extracted from two-dimensional localization fits to real-space fluorescence images of the qdot.  A linear fit is depicted as a grey solid line. \textit{Top inset:} Fluorescence image of a qdot displaced from the GNP. The latter also gives rise to very weak fluorescence. \textit{Bottom inset:} Fluorescne image of the coupled qdot-GNP system. Grey arrows indicate the data points corresponding to the insets. \textbf{e}, Measured fluorescence signal as a function of the lateral displacement during a coarse scan of the qdot across the GNP.  \textbf{f}, Fluorescence decay rate (inverse of the fluorescence lifetime) recorded simultaneously as (e) for each step of the coarse scan. \textbf{g}, Similar measurement as in \textbf{e} with a finer  sub-nanometer step size. \textbf{h}, Fluorescence decay rate (blue) and decay time (orange) as a function of the displacement. The grey dashed line represents the measurement decay rate limit dictated by the instrument response function.}
\label{fig:qdot}
\end{figure*}

The top horizontal axis in Fig.\,\ref{fig:vertical}c marks the voltage applied to the piezo element. Comparison of the two horizontal axes in Fig.\,\ref{fig:vertical}c indicates that the resulting displacement is linear with the applied voltage, corresponding to about \SI{2}{\nano\meter}/\SI{}{\volt}. The expected displacement without load amounts to \SI{380}{\nano\meter}/\SI{}{\volt}. We note that the aperture in the sample holder (see Fig.\,\ref{fig:setup}) allows for a slight bend of the lower substrate upon large pressure. This acts as a magnification in the piezo step resolution, facilitating a smooth gap reduction. In addition, the combination of the piezo actuator and the glass handle acts as a stiff spring mechanism, which reduces the sensitivity to vibrations. The quantitative details of this mechanical system depend on the specifics of the choices of the substrates and holders \cite{gulati201145}.

\section{Controlled enhancement of fluorescence from a single qdot coupled to a gold nanoparticle}
Modification of fluorescence in the near field of plasmonic nanostructures has continuously fascinated scientists since the early 1980s \cite{weitz1983enhancement, ruppin1982decay,chew1987transition}. A controlled and routine realization of this simple-seeming idea in the laboratory, however, continues to be elusive because it requires high degree of control in the shape, size, and material of the metallic nanostructure, nanometer precision in placement of the nanostructure with respect to an emitter, good control of the orientation of the emitter's dipole with respect to the nanostructure, and well-defined polarization of the excitation and illumination optical fields \cite{kuhn2008modification}. Various attempts have addressed some of these issue using ensemble statistical \cite{kinkhabwala2009large, chikkaraddy2016single, santhosh2016vacuum,leng2018strong}. Nearly two decades ago, we introduced a simple idea for performing controlled single-emitter studies: a gold nanoparticle was placed at the end of a glass tip to act as a nano-antenna, which could be positioned in all three dimensions with nanometer precision by using the machinery of a SNOM device \cite{kalkbrenner2001single,kuhn2006enhancement, anger2006enhancement}. While the quantitative control in this approach has attracted some attention \cite{matsuzaki2017strong, gross2018near,singh2018nanoscale,park2019tip}, its widespread use has been hampered by the experimental complexity that accompanies single-emitter SPM studies. We now demonstrate that PROscan can achieve similar results in a more robust arrangement in a concrete example, where a single qdots is coupled to a plasmonic nanoparticle.

\begin{figure*}[ht!]
	\centering
	\includegraphics[width=\textwidth]{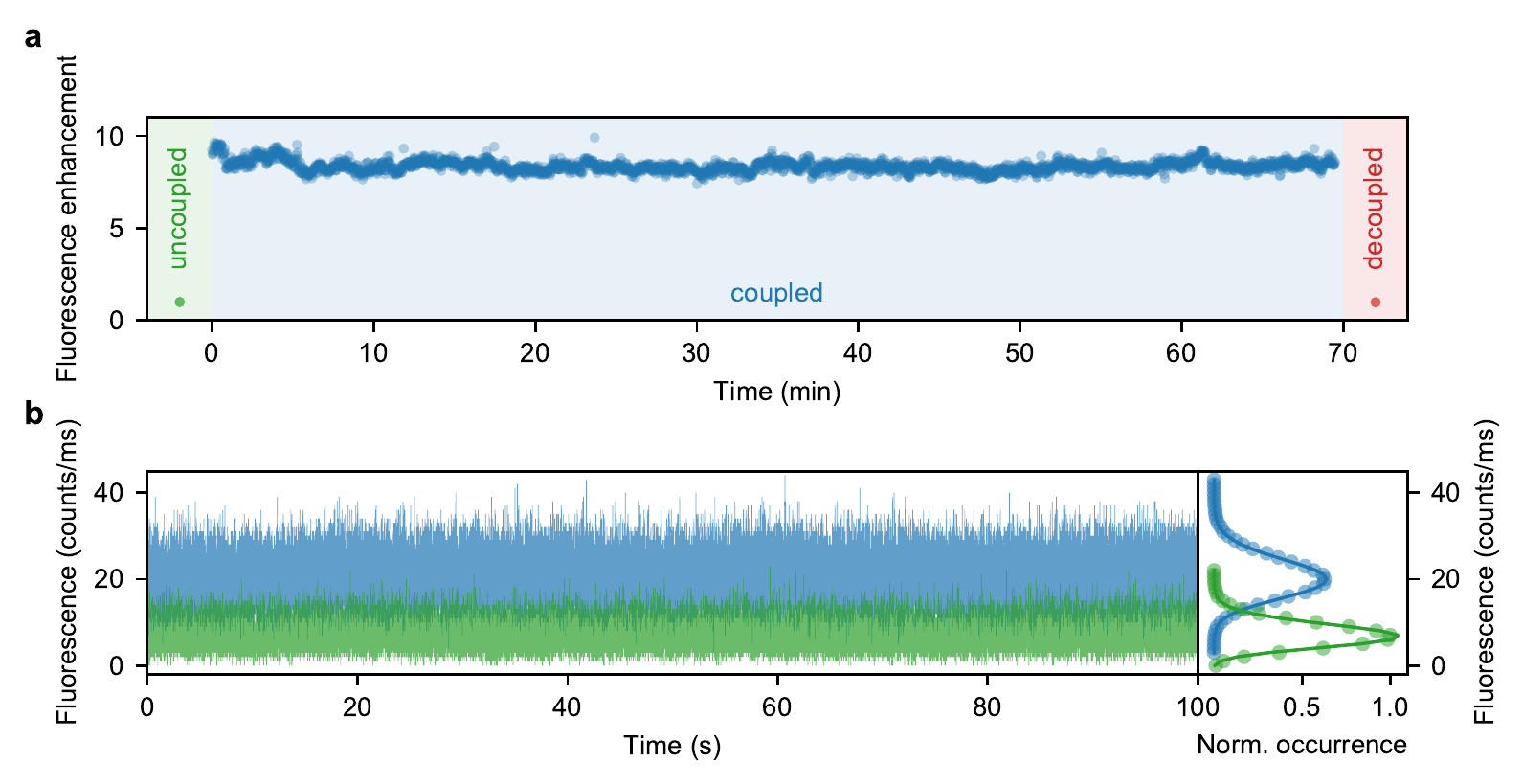}
	\caption{\textbf{a}, Fluorescence enhancement of the single qdot coupled to a single gold nanoparticle over the course of  \SI{70}{\minute} (blue). The green and red data points depict the fluorescence enhancement before and after the coupling event, respectiveley. \textbf{b}, \textit{Left}: Collected emission from an uncoupled (green) and coupled (blue) qdot presented in 1\,ms bins over the course of \SI{100}{\second }. \textit{Right}:  Histograms of the qdot fluorescence intensity in the coupled and uncoupled states. Solid curves show that the histograms fit Poissonian distributions very well.}
	\label{fig:Stability}
\end{figure*}

As depicted in Fig.\,\ref{fig:qdot}a, we placed qdots on the upper substrate and GNPs on the bottom substrate. The system was excited with a pulsed laser at a wavelength of \SI{532} {\nano\meter} and a repetition rate of \SI{4}{\mega\hertz}. For pulsed excitation well below saturation, the change of the fluorescence intensity is mainly expressed as the product of the modification factors in the excitation rate, quantum efficiency and collection efficiency \cite{matsuzaki2021quantum}. In our case, the latter is not a decisive factor since we collect the emitted light very efficiently with a high-NA microscope objective.

In Fig.\,\ref{fig:qdot}b, we show the emission spectra of a single qdot before (brown, center wavelength $\sim$652\,nm) and after coupling to a GNP antenna (blue, center wavelength $\sim$643\,nm), revealing an eleven-fold fluorescence enhancement. 
We note a blue-shifted shoulder in the emission spectrum, which can be attributed to a charged exciton or trion emission \cite{orfield2016quantum}. 
Furthermore, we observe a coupling-induced blue shift in the main emission peak. This effect is due to the detuning of the GNP plasmon resonance (peaks at \SI{560}{\nano\meter}) with respect to the emission spectrum of the qdot. 

We also recorded the fluorescence decay curve of the same qdot with and without the presence of the GNP. The results are shown by symbols in Fig.\,\ref{fig:qdot}c. The photophysics of the nonblinking qdots used in our work \cite{orfield2016quantum} allows for efficient access to bi-exciton emission. Indeed, we showed in a previous work that a gold nano-antenna can enhance radiative decay in both the exciton and bi-exciton channels of these qdots \cite{matsuzaki2017strong}. As expected, the decay curve of the uncoupled qdot displays a clear bi-exponential behavior, with fast ($\tau_1$) and slow ($\tau_2$) components, corresponding to the bi-exciton and exciton decay processes, respectively. The decay curve obtained for the coupled qdot reports a very fast drop of fluorescence by three orders of magnitude. By fitting this rapid decay by a bi-exponential function, we find that $\tau_1$ is reduced from $\SI{2.1\pm0.1}{\nano\second}$ to $\SI{0.4\pm0.1}{\nano\second}$, and $\tau_2$ is shortened from $\SI{29.4\pm0.8}{\nano\second}$ to $\SI{1.0\pm0.1}{\nano\second}$. 

Both measurements in Fig.\,\ref{fig:qdot}c were recorded as a part of a line scan, in which we positioned a qdot across the GNP. When increasing the piezo voltage in the $x$-direction, the initially unperturbed qdot couples to the GNP and is then decoupled again. The inset in the upper left corner of Fig.\,\ref{fig:qdot}d displays a fluorescence image of the qdot and the GNP (gold nanoparticles typically have a weak fluorescence signal \cite{he2008nonbleaching}), which are still well separated and distinguishable. We extract the positions of the GNP and the qdot by fitting two-dimensional Gaussian functions to the image. In this fashion, we deduce the distance between the qdot and the stationary GNP for each frame and plot it against the applied piezo voltage. The outcome is shown in Fig.\,\ref{fig:qdot}d, where the color of the symbols reflects the measured fluorescence signal. We note that lack of the knowledge of the dipole orientation and the position-dependence of the antenna effect on its radiation pattern prevent one from an accurate localization of the qdot very close to the GNP (see lower right inset of Fig.\,\ref{fig:qdot}d). To circumvent this problem, we fit the curve before and after the coupling event with a linear function and extrapolate the data points. 

Figure\,\ref{fig:qdot}e shows the integrated fluorescence signal of the qdot as a function of its distance to the GNP, normalized to the fluorescence of the uncoupled qdot. Furthermore, we plot the measured slow decay rate (exciton) in Fig.\,\ref{fig:qdot}f for each step and observe an initial increase from \SI{0.03}{\per \nano\second} for the uncoupled to \SI{1}{\per\nano\second} for the maximally coupled qdot in this scan. The asymmetric line profile might be caused by the tilt of the qdot dipole moment \cite{kuhn2008modification,matsuzaki2021quantum}. 

To examine the near-field interaction more closely, we repeated the measurements with a finer spatial resolution. Therefore, we again approached the qdot to the GNP until we observed a change of the fluorescence intensity at an approximate separation of $\sim\SI{70}{\nano\meter}$. Then, we laterally reduced the distance with a finer step size and recorded the fluorescence enhancement and temporal decay as presented in Figs.\,\ref{fig:qdot}g and \ref{fig:qdot}h, respectively. Acquiring independent information about step size is now more challenging because the distortion of the emission pattern in the near field of the GNP prevents one from measuring the qdot position using localization microscopy. Furthermore, because the step size per applied unit of voltage is not constant for different scan conditions, we cannot use the calibration obtained from Fig.\,\ref{fig:qdot}d. Thus, to estimate the scanned distance, we set the full width at half-maximum of the measured fluorescence signal of the fine scan in Fig.\,\ref{fig:qdot}g equal to that measured for the coarse scan presented in Fig.\,\ref{fig:qdot}d. 

The smooth profile of the fluorescence signal in Fig.\,\ref{fig:qdot}g demonstrates the ability of PROscan to explore the near field of the GNP with nanometer precision. One also notices a very sharp feature at about $x=15$\,nm. Given its well-defined rise and fall over a period longer than 10\,s (integration time per point was about 2\,s), we believe this is not an artifact. We attribute this event to the presence of a local sharp protrusion in the GNP. We also notice a rapid change in the fluorescence lifetime (right axis) and fluorescence decay rate (left axis) at $x=15$\,nm (see Fig.\,\ref{fig:qdot}h). The lifetime starts at about 3\,ns and falls, first softly and then very quickly, as the qdots approaches the GNP. The flat region at smaller distances denotes the finite instrument response time of our detector marked by the dashed grey line. Such a rapid shortening of the fluorescence lifetime is expected and is mostly due to the onset of nonradiative decay channels very close to gold \cite{kuhn2008modification,matsuzaki2021quantum}.


 
\section{Stability studies}
A decisive advantage of PROscan in comparison with conventional SPM platforms is the ability to engage in deep near field interactions with unprecedented mechanical stability. To demonstrate this feature, we monitored the fluorescence spectra of single qdots coupled to individual GNPs over more than one hour. In Fig.\,\ref{fig:Stability}a, we showcase the extraordinary stability of the PROscan device. The fluorescence enhancement is a particularly good measure for the system's stability because nanometer displacements between the nanoantenna and the qdot can drastically alter the signal. We attribute small fluctuations of the measured intensity to charging events of the qdot. To verify that both the qdot and the GNP are still placed on separate substrates during the long observation time (i.e., we did not accidentally pick one up), we disengaged the upper substrate and confirmed that the qdot fluorescence was reduced to its uncoupled value. The fluorescence of the qdot before coupling and after decoupling are marked by green and red colors, respectively, in Fig.\,\ref{fig:Stability}a.

In Fig.\,\ref{fig:Stability}b, we also plot the emitted fluorescence of a coupled qdot-GNP system (blue) recorded with higher temporal resolution with an APD. We observe that the emitted intensity binned in one millisecond follows a Poissonian distribution, similar to the emission trace of the same uncoupled non-blinking qdot (green). Hence, we conclude that there were no vibrations or oscillations that could have broadened the distribution.

\section{Conclusions}
Over the years, there have been many reports on the realization of measurement setups and geometries with inherent stability against mechanical vibrations, e.g., to investigate electrical and thermal conductivity of single molecules and atoms \cite{moreland1983squeezable,moreland1984electromagnetic,van1996adjustable,agrait2003quantum,vadai2013plasmon,popp2019ultra,gruber2020resonant}. These structures reach sub-nanometer precision in one dimension, but they have not demonstrated lateral scanning and positioning of a nanoprobe against a sample. We have shown that PROscan provides a sufficiently fine scannability to map the near-field coupling of a semiconductor qdot to a gold nanoantenna. We reported a substantial fluorescence enhancement during a horizontal scan accompanied by a reduction of the measured decay time. In particular, we used this coupled system to demonstrate the remarkable mechanical stability of this passive device over a period longer than one hour.

The PROscan apparatus is a powerful and yet simple alternative to conventional tip-based scanning probe techniques for executing a variety of nanoscopic optical, electrical and thermoelectrical measurements with nanometer resolution without a feedback mechanism. The tip-less design of the approach holds promise for incorporating planar scanning devices \cite{ernst2019planar}, scanning superconducting interference devices \cite{kirtley1995high}, scanning electron transistors \cite{yoo1997scanning}, and nitrogen vacancy centers in diamond \cite{maletinsky2012robust}. 

\section{Acknowledgements}
The authors thank Jennifer Hollingsworth and Matthew R. Buck for providing the quantum dots and Jan Renger for the SEM and helium ion microscopy images.
They acknowledge financial support by the Max Planck Society and by the QuantERA project RouTe through the Federal Ministry of Educationand Research (BMBF) (13N14839). This project has also received funding from the European Union’s Horizon 2020 research and innovation programme under the Marie Sk\l odowska Curie Grant Agreement No. 101025918.


\newpage

\medskip

\bibliography{Reference_PRoScan}

\end{document}